\documentclass[sigconf, authorversion, nonacm, screen]{acmart}

\usepackage{mathtools}
\usepackage{tikz}
\usepackage{enumitem}
\usepackage{listings}
\usetikzlibrary{automata}
\usetikzlibrary{arrows, arrows.meta, decorations.markings, patterns}
\usetikzlibrary{shapes, positioning, fit, backgrounds}
\usepackage{subfig}
\usepackage{pgfplots}
\pgfplotsset{compat=1.15,
	my linestyle/.style={
	thick,
	},
	my axis style/.style={
		xlabel style={
			font=\large\sffamily,
		},
		ylabel style={
			font=\large\sffamily,
		},
		grid,
	},		compat=newest}

\newtheorem{assum}{Assumption}
\newtheorem{defn}{Definition}

\newcommand{\real}{\mathbb{R}}

\definecolor{maroon}{cmyk}{0, 0.87, 0.68, 0.32}
\definecolor{halfgray}{gray}{0.55}
\definecolor{ipython_frame}{RGB}{207, 207, 207}
\definecolor{ipython_bg}{RGB}{247, 247, 247}
\definecolor{ipython_red}{RGB}{186, 33, 33}
\definecolor{ipython_green}{RGB}{0, 128, 0}
\definecolor{ipython_cyan}{RGB}{64, 128, 128}
\definecolor{ipython_purple}{RGB}{170, 34, 255}

\lstdefinelanguage{iPython}{
    morekeywords={access,and,break,class,continue,def,del,elif,else,except,exec,finally,for,from,global,if,import,in,is,lambda,not,or,pass,print,raise,return,try,while},%
    morekeywords=[2]{abs,all,any,basestring,bin,bool,bytearray,callable,chr,classmethod,cmp,compile,complex,delattr,dict,dir,divmod,enumerate,eval,execfile,file,filter,float,format,frozenset,getattr,globals,hasattr,hash,help,hex,id,input,int,isinstance,issubclass,iter,len,list,locals,long,map,max,memoryview,min,next,object,oct,open,ord,pow,property,range,raw_input,reduce,reload,repr,reversed,round,set,setattr,slice,sorted,staticmethod,str,sum,super,tuple,type,unichr,unicode,vars,xrange,zip,apply,buffer,coerce,intern},%
    sensitive=true,%
    morecomment=[l]\#,%
    morestring=[b]',%
    morestring=[b]",%
    morestring=[s]{'''}{'''},
    morestring=[s]{"""}{"""},
    morestring=[s]{r'}{'},
    morestring=[s]{r"}{"},%
    morestring=[s]{r'''}{'''},%
    morestring=[s]{r"""}{"""},%
    morestring=[s]{u'}{'},
    morestring=[s]{u"}{"},%
    morestring=[s]{u'''}{'''},%
    morestring=[s]{u"""}{"""},%
    literate=
    {á}{{\'a}}1 {é}{{\'e}}1 {í}{{\'i}}1 {ó}{{\'o}}1 {ú}{{\'u}}1
    {Á}{{\'A}}1 {É}{{\'E}}1 {Í}{{\'I}}1 {Ó}{{\'O}}1 {Ú}{{\'U}}1
    {à}{{\`a}}1 {è}{{\`e}}1 {ì}{{\`i}}1 {ò}{{\`o}}1 {ù}{{\`u}}1
    {À}{{\`A}}1 {È}{{\'E}}1 {Ì}{{\`I}}1 {Ò}{{\`O}}1 {Ù}{{\`U}}1
    {ä}{{\"a}}1 {ë}{{\"e}}1 {ï}{{\"i}}1 {ö}{{\"o}}1 {ü}{{\"u}}1
    {Ä}{{\"A}}1 {Ë}{{\"E}}1 {Ï}{{\"I}}1 {Ö}{{\"O}}1 {Ü}{{\"U}}1
    {â}{{\^a}}1 {ê}{{\^e}}1 {î}{{\^i}}1 {ô}{{\^o}}1 {û}{{\^u}}1
    {Â}{{\^A}}1 {Ê}{{\^E}}1 {Î}{{\^I}}1 {Ô}{{\^O}}1 {Û}{{\^U}}1
    {œ}{{\oe}}1 {Œ}{{\OE}}1 {æ}{{\ae}}1 {Æ}{{\AE}}1 {ß}{{\ss}}1
    {ç}{{\c c}}1 {Ç}{{\c C}}1 {ø}{{\o}}1 {å}{{\r a}}1 {Å}{{\r A}}1
    {€}{{\EUR}}1 {£}{{\pounds}}1
    {^}{{{\color{ipython_purple}\^{}}}}1
    {=}{{{\color{ipython_purple}=}}}1
    {+}{{{\color{ipython_purple}+}}}1
    {*}{{{\color{ipython_purple}$^\ast$}}}1
    {/}{{{\color{ipython_purple}/}}}1
    {+=}{{{+=}}}1
    {-=}{{{-=}}}1
    {*=}{{{$^\ast$=}}}1
    {/=}{{{/=}}}1,
    literate=
    *{-}{{{\color{ipython_purple}-}}}1
     {?}{{{\color{ipython_purple}?}}}1,
    identifierstyle=\color{black}\ttfamily,
    commentstyle=\color{ipython_cyan}\ttfamily,
    stringstyle=\color{ipython_red}\ttfamily,
    keepspaces=true,
    showspaces=false,
    showstringspaces=false,
    numbers=left,
    numberstyle=\tiny\color{halfgray},
    frame=lines,
    framesep=2mm,
    keywordstyle=\color{ipython_green}\ttfamily,
}
\newcommand*{\tran}{^{\mkern-1.5mu\mathsf{T}}\!}  

\def\d{\ensuremath{\mathrm{d}}}

\def\norm[#1]{\left|#1\right|}
\def\shortnorm[#1]{|#1|}

\def\Xs{\mathcal{X}}
\def\Ys{\mathcal{Y}}
\def\Us{\mathcal{U}}
\def\Ws{\mathcal{W}}

\def\R{\mathbb{R}}

\def\Q{\mathbb{Q}}

\def\Is{\mathcal{I}}

\def\Rs{\mathcal{R}}

\def\Ss{\mathcal{S}}

\def\Qs{\mathcal{Q}}

\def\Es{\mathcal{E}}  
\def\Cs{\mathcal{C}}  
\def\Us{\mathcal{U}}  
\def\Bb{\mathcal{B}} 

\def\xv{\boldsymbol{x}}

\def\zetav{\boldsymbol{\zeta}}

\def\Am{\boldsymbol{A}}
\def\Bm{\boldsymbol{B}}

\def\I{{\normalfont\textbf{I}}}

\def\Mm{\boldsymbol{M}}
\def\Nm{\boldsymbol{N}}

\def\Pm{\boldsymbol{P}}
\def\Qm{\boldsymbol{Q}}

\def\Km{\boldsymbol{K}}

\def\Pl{\Pm_{\mathrm{Lyap}}}
\def\Ql{\Qm_{\mathrm{Lyap}}}

\def\e{\mathrm{e}}

\def\edge#1{\xrightarrow{#1}}

\AtBeginDocument{%
  \providecommand\BibTeX{{%
    \normalfont B\kern-0.5em{\scshape i\kern-0.25em b}\kern-0.8em\TeX}}}

\copyrightyear{2022} 
\acmYear{2022} 
\setcopyright{rightsretained} 
\acmConference[HSCC '22]{25th ACM International Conference on Hybrid Systems: Computation and Control}{May 4--6, 2022}{Milan, Italy}
\acmBooktitle{25th ACM International Conference on Hybrid Systems: Computation and Control (HSCC '22), May 4--6, 2022, Milan, Italy}\acmDOI{10.1145/3501710.3519523}
\acmISBN{978-1-4503-9196-2/22/05}

\begin{document}

\title{ETCetera: beyond Event-Triggered Control}


\author{Giannis Delimpaltadakis}
\email{i.delimpaltadakis@tudelft.nl}
\affiliation{%
  \institution{Delft University of Technology}
  \city{Delft}
  \country{The Netherlands}}
\author{Gabriel de A. Gleizer}
\email{g.gleizer@tudelft.nl}
\affiliation{%
  \institution{Delft University of Technology}
  \city{Delft}
  \country{The Netherlands}}
\author{Ivo van Straalen}
\email{ivo65@live.nl}
\affiliation{%
  \institution{Delft University of Technology}
  \city{Delft}
  \country{The Netherlands}}
\author{Manuel Mazo Jr.}
\email{m.mazo@tudelft.nl}
\affiliation{%
  \institution{Delft University of Technology}
  \city{Delft}
  \country{The Netherlands}}



\begin{abstract}
 We present ETCetera, a Python library developed for the analysis and synthesis of the sampling behaviour of event triggered control (ETC) systems. In particular, the tool constructs abstractions of the sampling behaviour of given ETC systems, in the form of timed automata (TA) or finite-state transition systems (FSTSs). When the abstraction is an FSTS, ETCetera provides diverse manipulation tools for analysis of ETC's sampling performance, synthesis of communication traffic schedulers (when networks shared by multiple ETC loops are considered), and optimization of sampling strategies. Additionally, the TA models may be exported to UPPAAL for analysis and synthesis of schedulers. Several examples of the tool's application for analysis and synthesis problems with different types of dynamics and event-triggered implementations are provided.
\end{abstract}



\maketitle

\section{Introduction}

Since the seminal works of {\AA}str\"om \cite{astrom2002comparison} and {\AA}rzen~\cite{arzen1999}, and in particular since the work of Tabuada~\cite{tabuada2007event}, there has been a surge of interest in the use of event-based controller update techniques in order to reduce the resource consumption of (networked) control systems. Despite the myriad of proposals made in the literature, very little attention has been given, until recently, to the quantification of the benefits of such proposals. As such, little was known about the sampling patterns arising from so called event-triggered control (ETC) systems, what is their efficiency (e.g. what is the average inter-sample time), or how hard are they to schedule.

In the past decade this gap in the literature was addressed in a number of papers. Both analytical \cite{postoyan_interevent,tallapragada} and computational approaches \cite{gleizer2020scalable,gleizer2020towards,gleizer2021cdc,gleizer2021computing,gleizer2021hscc,delimpa_cdc_homog,delimpaltadakis2021traffic,delimpaltadakis2021stochastic,arman_formal_etc,mazo2018book} have been pursued to study ETC's sampling behaviour. 
Particularly, the computational approach builds on the notion of abstractions. These abstractions distil the sampling-times' dynamics from ETC's hybrid dynamical model
, and serve as a computationally tractable object which may be employed to address a wide range of verification and synthesis questions on ETC's sampling patterns. For their construction,  the state-space of the control system is partitioned in an informed way. The resulting abstraction is either a timed-automaton (TA), or in the case of periodic ETC (PETC) a finite transition system (FTS), capturing the sampling times' dynamics. We call these objects \emph{traffic models}, evoking how they model the communication traffic (of measurements and actuation commands) generated by the control system.

Once such traffic models are available they can be employed for two main purposes: \emph{analysis} of metrics such as average inter-sample times measuring the efficiency of an ETC design \cite{gleizer2021computing,gleizer2021hscc}; or for \emph{synthesis} of scheduling strategies \cite{mazo2018book}, or optimized sampling strategies \cite{gleizer2021cdc}. Synthesizing scheduling strategies is a necessary capability when implementing event triggered control systems sharing a communication medium. The aim of the schedulers synthesized is to prevent multiple control-loops (or other real-time tasks) requesting a channel access simultaneously, which would lead to communication collisions and potentially to control instability. One may achieve such scheduling goal by enabling earlier transmissions than dictated by the ETC mechanism, or even by allowing switching between different triggering conditions. In a similar way, the traffic models can be employed to optimize sampling strategies by enforcing, at times, earlier sampling of the control loop than that dictated by ETC. The rationale in this optimization is that while one is generally interested in minimizing the \emph{cumulative} or \emph{average} amount of communications employed, ETC approaches are greedy in trying to optimize such goals, i.e. they try to enlarge as much as possible each individual inter-sample time interval. In general, greedy optimization approaches are suboptimal, and as such early triggering can be used as a control variable to optimize beyond what the greedy ETC strategies can attain.

All of these developments: model construction, analysis, and synthesis of schedulers or sampling strategies, are automated techniques that rely on heavy computation. In the present paper we describe a toolbox that implements all these developments. The tool is named ETCetera and is available at \url{https://github.com/sync-lab/ETCetera}. It allows the user to:
\begin{enumerate}
    \item Construct traffic models of nonlinear ETC systems with arbitrary triggering conditions and disturbances;
    \item Construct specialized traffic models of Linear PETC systems with Quadratic triggering conditions;
    \item Export models to UPPAAL \cite{uppaal_stratego} for scheduler synthesis;
    \item Synthesize schedulers effectively for several PETC systems sharing a channel;
    \item Synthesize optimized sampling strategies; 
    \item Analyze PETC systems through the computation of traffic metrics;
\end{enumerate}


In what follows, we provide a brief introduction to event-triggered control and other needed preliminaries (Section \ref{sec:prelim}), describe the steps involved in the abstractions the tool constructs (Sections \ref{ssec:abstraction_general} and \ref{ssec:abstraction_linear}), present the approach implemented in ETCetera for PETC scheduler design (Section \ref{sec:scheduler synthesis}), and briefly introduce other synthesis and analysis problems also implemented in the tool (section \ref{ssec:quant}). Finally,
Section 4 provides the basic instructions of use of the tool, which are then illustrated on several examples.

\section{Preliminaries}\label{sec:prelim}

\subsection{Event-triggered control}
Consider a control system with state-feedback:
\begin{equation*}
    \dot{\zeta}(t) = f(\zeta(t),u(\zeta(t)), d(t))
\end{equation*}
where $t$ is time, $\zeta(t)\in\real^{n_\zeta}$ is the state, $u(t)\in\real^{n_u}$ is the control input, $d(t)\in\real^{n_d}$ is an unknown signal (e.g., a disturbance) and $f:\real^{n_\zeta}\times\real^{n_u}\times\real^{n_d}\to\real^{n_\zeta}$ is the vector field. In \textit{sample-and-hold} control implementations, the input is held constant between consecutive \textit{sampling times} $t_i,t_{i+1}$:
\begin{equation}\label{etc_sys}
    \dot{\zeta}(t) = f(\zeta(t),u(\zeta(t_i)), d(t)), \quad t\in[t_i,t_{i+1})
\end{equation}
Specifically, in \textit{continuous event-triggered control} (CETC), sampling times are defined as follows:
\begin{equation}\label{cetc_trig_cond}
    \begin{aligned}
&t_{i+1}=t_i + \min\Big\{\tau_{\max},\inf\{t>0:\phi(\zeta(t),\zeta(t_i))\geq0\}\Big\}
    \end{aligned}
\end{equation}
where $t_0=0$, \eqref{cetc_trig_cond} is called the \textit{triggering condition}, $\phi(\cdot,\cdot)$ is called the \textit{triggering function}, and the difference $t_{i+1}-t_i$ is called the \textit{inter-sample time}. The constant $\tau_{\max}$ is a forced upper bound on inter-sample times, and it prevents the system from running open-loop for an indefinite amount of time. Between two consecutive sampling times $t_i$ and $t_{i+1}$, the triggering function starts from a negative value $\phi(\zeta(t_i),\zeta(t_i))<0$, and remains negative until it becomes zero at $t^-_{i+1}$; at $t^+_{i+1}$, the state $\zeta(t_{i+1})$ is sampled again, the control input is updated $u(\zeta(t_{i+1}))$, the triggering function resets to a negative value, and the process is repeated again. As the inter-sample time depends solely on the initial condition $\zeta(t_i)$ and the unknown signal $d(t)$, we denote it by:
\begin{equation}\label{inter-sample_time}
\begin{aligned}
    \tau_r(x) := \min\Big\{&\tau_{\max},\\&\inf\{t>0:\phi(\zeta(t),\zeta(t_i)), \zeta(t_i)=x, d(t) = r(t)\}\Big\}    
\end{aligned}
\end{equation}
and when $d(t)=0$ (e.g., unperturbed systems) we just write $\tau(x)$.

CETC owes its name to the fact that the triggering condition is checked continuously by the sensors, to detect \textit{events} (moments when the triggering function becomes zero) right when they happen. In contrast to this regime, in a special variant of ETC termed \textit{periodic ETC} (PETC), the triggering condition is only checked periodically in time. In such cases, the sampling times are defined as:
\begin{equation}\label{petc_trig_cond}
    \begin{aligned}
    &t_{i+1}=t_i + \min\Big\{k_{\max}h,\inf\{kh:k\in\mathbb{N},\phi(\zeta(kh),\zeta(t_i))\geq 0\}\Big\}
    \end{aligned}
\end{equation}
where $h$ is the \textit{checking period} and $k_{\max}\in\mathbb{N}$ again determines an upper bound on inter-sample times.
\subsection{Transition systems}

The ETC traffic models built by the ETCetera tool can be encapsulated as \emph{generalized transition systems}, following the definition from Tabuada in \cite{tabuada2009verification}:
	\begin{defn}[Transition system]\label{def:system} 
		A system $\Ss$ is a tuple $(\Xs,\Xs_0,\allowbreak \Us,\Es,\Ys,H)$ where:
		\begin{itemize}
			\item $\Xs$ is the set of states,
			\item $\Xs_0 \subseteq \Xs$ is the set of initial states,
			\item $\Us$ is the set of actions,
			\item $\Es \subseteq \Xs \times \Us \times \Xs$ is the set of edges (or transitions),
			\item $\Ys$ is the set of outputs, and
			\item $H: \Xs \to \Ys$ is the output map.
		\end{itemize}
	\end{defn}

A (transition) system is said to be \emph{finite} if $\Xs$ and $\Us$ are finite; this is the case of \emph{automata}, for which efficient tools exist for verification of properties and synthesis of controllers. A close relative, but infinite-state, for which such tools also exist is a \emph{timed automaton} \cite{alur1994ta}, in which a finite system is equipped with a finite set of \emph{clocks} $c_i$ whose values can only vary according to $\dot{c}_i = 1$ or be reset on edges. A transition on a system is denoted by $x\edge{u}x'$ meaning that $(x,u,x') \in \Es.$ An infinite \emph{run} of a system is a sequence $x_0\edge{u_0}x_1\edge{u_1}x_2...$, and the set of all possible runs of $\Ss$ is called its \emph{internal behavior set}. Every run maps into an \emph{external behavior}, or \emph{trace} $y_0u_0y_1u_1y_2...$ by taking $y_i = H(x_i)$. The set of all possible traces (or \emph{infinite external behavior set}) of $\Ss$ is denoted by $\Bb_\Ss$.

An \emph{abstraction} is a simpler version of a system preserving some of its properties. For verification, an abstraction needs to preserve the external behavior set $\Bb_\Ss$ (possibly including additional ``spurious'' behaviors), while for control synthesis it needs to, loosely writing, preserve the behaviors that can be \emph{enforced} by the controller through its action set. The associated formal notions are, respectively, \emph{(bi)simulations} and \emph{alternating (bi)simulations}. See \cite{tabuada2009verification} for the formal definitions and associated methods. In this work, ETC systems and their sampling behaviors are abstracted by finite transition systems or timed automata; in this way, the constructed abstractions are used either to verify properties (verification) of a given ETC system's behavior or to synthesize traffic schedulers (control synthesis) that avoid packet collisions in networks shared by multiple ETC loops.

\section{Theoretical Underpinnings}\label{sec:theory}

\subsection{Traffic abstractions of CETC systems with disturbances}\label{ssec:abstraction_general}

The ETCetera tool constructs abstractions that capture the sampling behavior of a given ETC system: \textit{traffic models}. In the general case, where the given system \eqref{etc_sys}-\eqref{cetc_trig_cond} may be a nonlinear CETC system with disturbances or uncertainties and the triggering function is of general form, the methods used are the ones developed in \cite{delimpa_cdc_homog,delimpaltadakis2021traffic}. For the construction of these abstractions, among some technical assumptions (see \cite{delimpaltadakis2021traffic}), we assume the following:
\begin{assum}\hfill
\begin{itemize}[leftmargin=1.5em]
    \item The system operates in a bounded hyper-rectangular state-space $X\subset\real^{n_\zeta}$ (parameter \lstinline[basicstyle=\ttfamily\small, language=iPython]{state_space_limits}).
    \item The unknown signal $d(t)$ is bounded by a hyper-rectangular domain $D\subset\real^{n_d}$ (parameter \lstinline[basicstyle=\ttfamily\small, language=iPython]{disturbance_limits}): $d(t)\in D$ for all $t$.
\end{itemize}
\end{assum}
The constructed abstraction is a transition system $\Ss=(\Xs,\Xs_0,\Es,\allowbreak\Ys,H)$ where:
\begin{itemize}
    \item $\Xs_0=\Xs:=\{\Rs_i:i\in\Is\}$, where $\Is$ is a finite set of indices (e.g., $\Is = \{1,2,\dots,q\}$), and $\bigcup_{i\in\Is}\Rs_i = X$; i.e., the sets $\Rs_i$ constitute a partition of the ETC system's state space. 
    \item $\Ys\subseteq 2^{\real^+}$ and $H(\Rs_i) = [\underline{\tau}_{\Rs_i}, \overline{\tau}_{\Rs_i}]$, where:
    \begin{equation}\label{ts_intervals}
    \begin{aligned}
        &\underline{\tau}_{\Rs_i}\leq \inf\limits_{x\in\Rs_i,\text{ } d(t)\in D}\tau_d(x)\\ 
        &\overline{\tau}_{\Rs_i}\geq \sup\limits_{x\in\Rs_i,\text{ } d(t)\in D}\tau_d(x)
    \end{aligned}
    \end{equation}
    \item $(\Rs_i,\Rs_j)\in\Es$, if $\exists$ $x\in\Rs_i$ and $t\in H(\Rs_i)$ s.t.~$\zeta(t;x)\in\Rs_j$.
\end{itemize}
We have omitted the set of actions $\Us$, since the constructed abstractions are autonomous and they are only meant to abstract the sampling behavior of the given system (for how these abstractions may later be endowed with actions, to control the system's sampling and schedule ETC network traffic, see Section \ref{sec:scheduler synthesis}). Each state of the abstraction $\Rs_i$ represents a region in the state space of the ETC system. Observe that the output of state $\Rs_i$ is an interval containing all possible inter-sample times that may be exhibited by the system if initialized in $\Rs_i$ (under any admissible realization of the unknown signal $d(t)$). Due to these observations and how the transitions have been defined, it can be seen that: for any sequence of inter-sample times $\tau_1,\tau_2,\dots$ that the ETC system \eqref{etc_sys}-\eqref{cetc_trig_cond} may exhibit, there is a corresponding trace of the abstraction $[\underline{\tau}_{\Rs_1}, \overline{\tau}_{\Rs_1}], [\underline{\tau}_{\Rs_2}, \overline{\tau}_{\Rs_2}], \dots$ such that $\tau_i\in [\underline{\tau}_{\Rs_i}, \overline{\tau}_{\Rs_i}]$ for all $i$. Thus, the abstraction captures the ETC system's sampling behavior. In fact, one can establish formally that the abstraction \textit{$\epsilon$-approximately simulates} the ETC system's sampling behavior (see \cite{arman_formal_etc,mazo2018book}). What is more, as elaborated in \cite{mazo2018book}, it is semantically equivalent to a \textit{timed automaton} (e.g., see \cite{alur1994ta}), where the intervals $[\underline{\tau}_{\Rs_i}, \overline{\tau}_{\Rs_i}]$ appear appropriately in the automaton's \textit{guards} and \textit{invariants}. As such, once the traffic model has been constructed, the numerous algorithms that have been developed for timed automata (e.g., \cite{uppaal_stratego})  may be used for verification and/or traffic scheduler synthesis.

As described in \cite{delimpaltadakis2021traffic}, the intervals $[\underline{\tau}_{\Rs_i}, \overline{\tau}_{\Rs_i}]$ are computed via an iterative algorithm combining a line-search and reachability analysis computational tools (dReach \cite{dreach} and Flow* \cite{flowstar}). Specifically, we run a line-search over the variables $\underline{\tau}_{\Rs_i}$ and $\overline{\tau}_{\Rs_i}$ until the following are verified by the reachability analysis tools:
\begin{equation}\label{intervals}
    \begin{aligned}
        &\bullet\phi(\zeta(t;x),x)<0, \quad \forall t\in[0,\underline{\tau}_{\Rs_i}],\forall x\in\Rs_i,\forall d(t)\in D\\
        &\bullet\phi(\zeta(\overline{\tau}_{\Rs_i};x),x)\geq0, \forall x\in\Rs_i,\forall d(t)\in D
    \end{aligned}
\end{equation}
Indeed, as shown in \cite{delimpa_cdc_homog,delimpaltadakis2021traffic}, the above conditions can be transformed to set-membership properties of the ETC system's trajectories, and thus checked by dReach or Flow*. As soon as the line-search finds parameters $\underline{\tau}_{\Rs_i}$ and $\overline{\tau}_{\Rs_i}$ for which \eqref{intervals} holds, the algorithm stops, as \eqref{intervals} implies \eqref{ts_intervals}. Observe that, in this way, the determined intervals are generally larger than the tightest possible intervals dictated by \eqref{ts_intervals}. Larger intervals imply more non-determinism in the abstraction (the abstraction approximates the actual sampling behavior more coarsely).

The transitions $\Es$ are also computed via reachability analysis, since the condition for transitions presented above may also be transformed to set-membership properties of the system's trajectories. Note that the computations performed by reachability analysis tools are not exact (e.g. the sets and the trajectories are overapproximated); this may lead to \textit{spurious transitions} (transitions not exhibited by the actual system's behavior), further contributing to the abstraction's non-determinism.

Moreover, the tool supports two different ways of partitioning the state space $X$ into regions $\Rs_i$: 

\textbf{Grid}: Creating a grid over $X$ via hyper-rectangles $\Rs_i$. The user has to specify parameters $(p_1,p_2,\dots,p_{n_{\zeta}})$ (\hspace{-1mm} \lstinline[basicstyle=\ttfamily\small, language=iPython]{grid_points_per_dim}), such that $X$ is partitioned into $p_1\times p_2\times\dots\times p_{n_{\zeta}}$ equal hyper-rectangles  $\Rs_i$.

\begin{figure}
		\centering
		\includegraphics[width=3in]{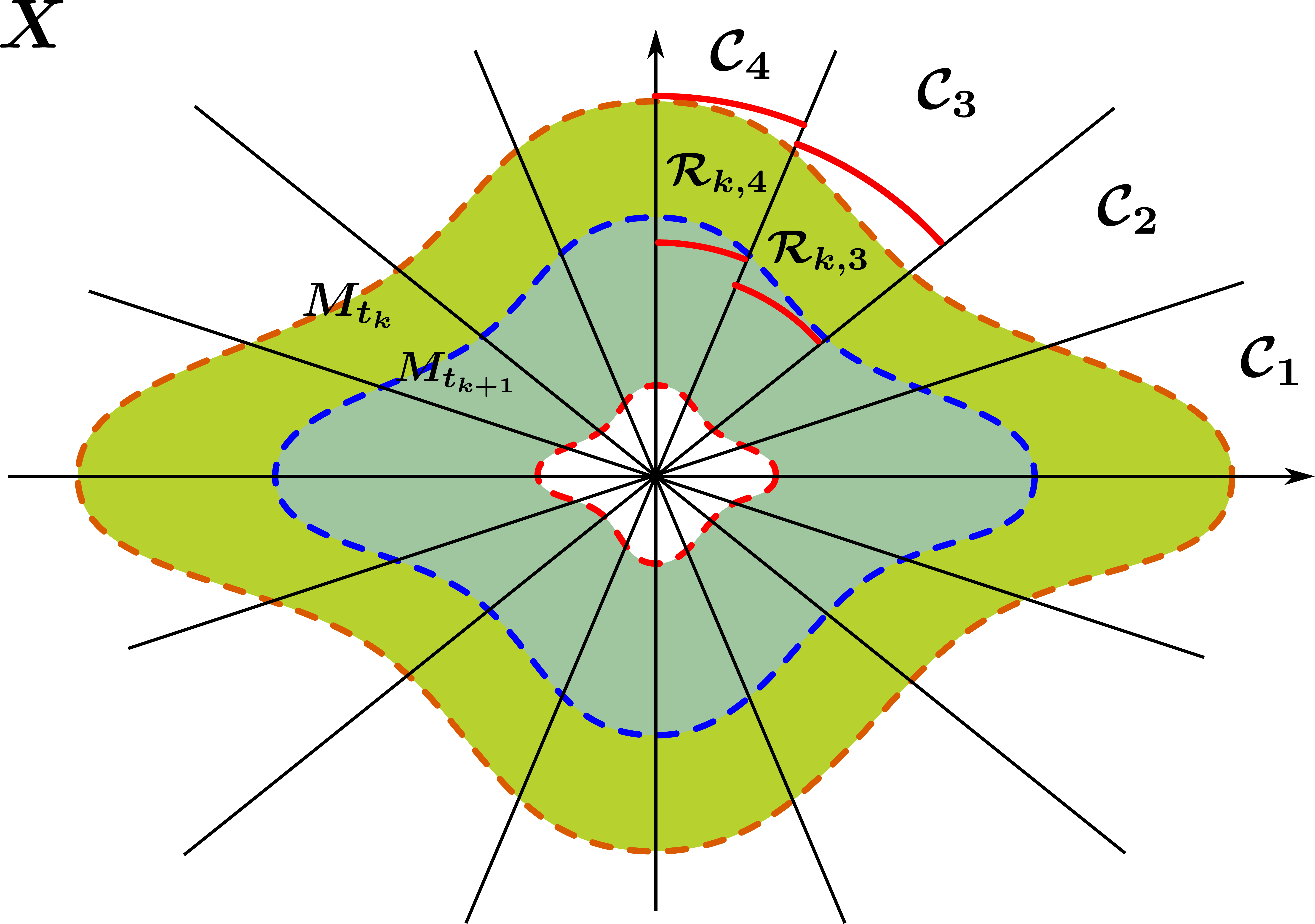}
		\caption{IM-based partitioning for a homogeneous CETC system. The rays passing through the origin determine the cones $\Cs_j$. The dashed lines are consecutive approximations of IMs $M_{t_k}$ and $M_{t_{k+1}}$. The regions $\Rs_{k,j}$ are delimited by consecutive approximations of IMs and cones (hence the indexing). The circular segments in cones $\Cs_3$ and $\Cs_4$ delimit the ball segments that overapproximate regions $\Rs_{k,3}$ and $\Rs_{k,4}$.}
		\label{fig:discr_homogeneous}
\end{figure}

\begin{figure}
		\centering
		\includegraphics[width=3in]{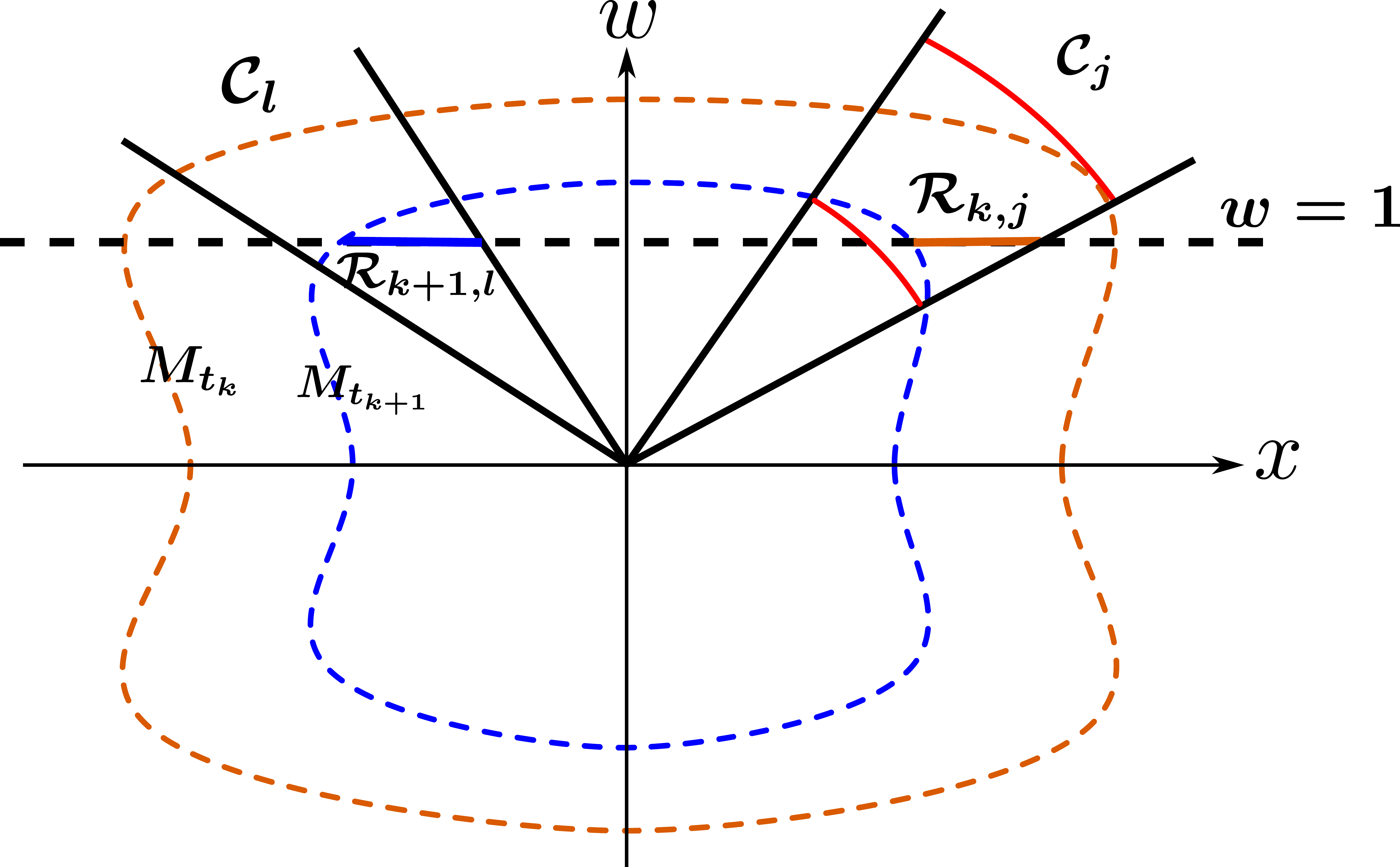}
		\caption{IM-based partitioning for a non-homogeneous CETC system, which has been embedded in $\real^{n_\zeta + 1}$. The mapped state-space $\tilde{X}$ is a subset of the $w=1$-plane. The rays passing through the origin determine the polyhedral cones $\Cs_j$. The dashed curves are consecutive approximations of IMs $M_{t_k}$ and $M_{t_{k+1}}$. The regions $\Rs_{k,j}$ are intersections between cones $\Cs_j$, the sets between consecutive approximations of isochronous manifolds $M_{t_k}$ and $M_{t_{k+1}}$ and the set $\tilde{X}$ (hence the indexing). The circular segments in cone $\Cs_j$ delimit the ball segment, say $\mathcal{B}_{k,j}$, which defines an overapproximation of $\Rs_{k,j}$ as follows: $\Rs_{k,j}\subseteq\mathcal{B}_{k,j}\cap\tilde{X}$.}
		\label{fig:discr_nonhomogeneous}
\end{figure}

\textbf{IM-based}: Partitioning $X$ via approximations of \textit{isochronous manifolds}\footnote{Isochronous manifolds $M_{t_k}$ are sets of points in the system's state-space that correspond to the same inter-sample time $t_k$; i.e., the system initialized anywhere in $M_{t_k}$ exhibits the same inter-sample time $t_k$ (under the same realization of the unknown signal $d(t)$). For more information, see \cite{delimpa_tac,delimpa_tcns}.} (IMs) and cones, and obtaining $\Rs_i$ as the sets that are delimited by these approximations and cones. In particular, due to theoretical subtleties, the state space partitioning is done slightly differently, depending on the dynamics: 
\begin{itemize}[leftmargin=*]
    \item \textit{Homogeneous Systems with $d(t)=0$}: When the given system is \textit{homogeneous} (for the definition, see \cite{delimpa_tac}) and $d(t)=0$, the process is as follows. To compute approximations of isochronous manifolds $M_{t_k}$, the user has to specify the times $(t_1,\dots,t_k,\dots,t_q)$ (parameter \lstinline[basicstyle=\ttfamily\small, language=iPython]{manifolds_times}, in ascending order) to which the manifolds correspond. To determine the cones $\Cs_j$, we follow a similar process to the \textit{isotropic covering} of \cite{fiter2012state}, in which each $i$-th angular spherical coordinate of the state-space is partitioned into $a_i$ equal angles. Thus, the user has to specify the parameters $(a_1,\dots,a_j,\dots,a_{n_{\zeta}-1})$ (\lstinline[basicstyle=\ttfamily\small, language=iPython]{angles_discretization}), to obtain at most\footnote{Some of the constructed cones are discarded, because they are degenerate (Lebesgue measure-zero). Nonetheless, the union of all remaining non-degenerate cones constitutes a partition of $\real^{n_{\zeta}}$.} $a_1\times a_2 \times \dots a_{n_{\zeta}-1}$ cones $\Cs_j$. Finally, the regions $\Rs_i$ are obtained as intersections between cones $\Cs_j$, sets between consecutive approximations of isochronous manifolds $M_{t_k}$ and $M_{t_{k+1}}$ and the set $X$. Moreover, since the regions $\Rs_i$ admit a representation that is complex for the employed reachability analysis tools, we overapproximate them via ball-segments contained inside the cones $\Cs_j$. Then, dReach is used for reachability analysis, as it can handle ball-segments efficiently. An illustration of IM-based partitioning for homogeneous systems is shown in Fig. \ref{fig:discr_homogeneous}.
    \item \textit{Non-homogeneous Systems with $d(t)=0$}: When the system is \textit{non-homogeneous} and $d(t)=0$, it is lifted to $\real^{n_\zeta+1}$ by adding a dummy variable $w$, where it is rendered homogeneous. The state-space $X$ of the original system is now mapped to the set $\tilde{X}=\{(x,1)\in\real^{n_{\zeta}+1}:x\in X\}$. The user has to specify parameters $(p_1,p_2,\dots,p_{n_{\zeta}})$ (\lstinline[basicstyle=\ttfamily\small, language=iPython]{grid_points_per_dim}), such that $\tilde{X}$ is partitioned into $p_1\times p_2\times\dots\times p_{n_{\zeta}}$ equal hyper-rectangles. Each of these hyper-rectangles determines a polyhedral cone $\Cs_j$ pointed at the origin, with its extreme vertices being the vertices of the hyper-rectangle. Furthermore, as in the previous case, the times $(t_1,\dots,t_k,\dots,t_q)$ (parameter \lstinline[basicstyle=\ttfamily\small, language=iPython]{manifolds_times}, in ascending order) have to be given, to derive approximations of isochronous manifolds $M_{t_k}$. Finally, the regions $\Rs_i$ are obtained as intersections between cones $\Cs_j$, the sets between consecutive approximations of isochronous manifolds $M_{t_k}$ and $M_{t_{k+1}}$ and the set $\tilde{X}$. As above, the regions $\Rs_i$ are overapproximated by ball-segments contained in $\Cs_j$, and reachability analysis is carried out via dReach. An illustration of IM-based partitioning for non-homogeneous systems is shown in Fig. \ref{fig:discr_nonhomogeneous}.   
    \item $d(t)\neq0$: In this case, even though the theory developed in \cite{delimpaltadakis2021traffic} does address it, IM-based partitioning is not yet supported by ETCetera, due to the fact that dReach cannot handle disturbances (and Flow* cannot handle ball-segments). We are working towards addressing this issue in future versions of the tool.
\end{itemize}
Finally, the approximations of IMs are computed with the iterative algorithm described in \cite{delimpa_tac}, which employs the SMT solver dReal \cite{dreal}. The user may specify dReal's precision parameter $\delta$ (\lstinline[basicstyle=\ttfamily\small, language=iPython]{precision_deltas}, with default value $10^{-7}$), which affects the accuracy of approximation; the smaller $\delta$ is, the more accurate the approximations are, but the computational load becomes larger.

Generally, grid-partitioning is more computationally efficient w.r.t. the computational load of creating the abstraction. On the other hand, IM-based partitioning generally provides tighter intervals $[\underline{\tau}_{\Rs_i}, \overline{\tau}_{\Rs_i}]$, thus reducing one of the abstraction's sources of non-determinism. However, the current implementation of the tool does not check if the given \lstinline[basicstyle=\ttfamily\small, language=iPython]{manifolds_times} result into approximations of isochronous manifolds which cover the whole state-space of interest $X$; thus, in IM-based partitioning, the user has to inspect the end-result to verify that the state-space is indeed covered by the regions $\Rs_i$. 
Finally, we note that even when grid-partitioning is employed, the tool uses approximations of isochronous manifolds to derive estimates on timing lower bounds $\underline{\tau}_{\Rs_i}$ and use them as a starting-point for the line-search algorithms, thus speeding up computations. For more information on the abstractions' construction and on approximations of isochronous manifolds the reader is referred to \cite{delimpa_cdc_homog,delimpaltadakis2021traffic} and \cite{delimpa_tac,delimpa_tcns}, respectively.

\subsection{Traffic abstractions of linear PETC systems}\label{ssec:abstraction_linear} 

In the case of unperturbed linear systems under PETC, the construction of abstractions can be performed with simpler computations, provide exact simulations (thanks to the periodic nature of condition checking), and in general the abstractions are tighter than when employing the general scheme described in the previous section.

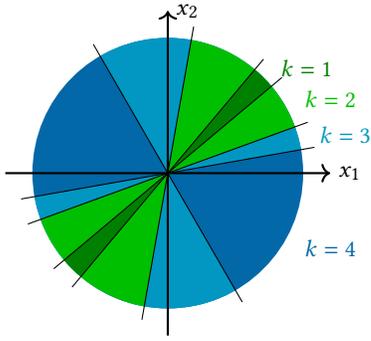
\begin{figure}
    \centering
    \colorlet{dgreen}{green!75!black}
\colorlet{ddgreen}{green!50!black}
\colorlet{dcyan}{cyan!75!black}
\colorlet{aqua}{dcyan!50!blue}
\begin{tikzpicture}[scale=1.8]
\fill[aqua] (0,0) circle [radius=1];
\fill[fill=dcyan] (0,0) -- (10:1) arc (10:120:1) -- (0,0);
\fill[fill=dgreen] (0,0) -- (20:1) arc (20:80:1) -- (0,0);
\fill[fill=ddgreen] (0,0) -- (40:1) arc (40:50:1) -- (0,0);
\draw (0,0) -- (10:1.1);
\draw (0,0) -- (120:1.1); 
\draw (0,0) -- (20:1.1);
\draw (0,0) -- (80:1.1) ;
\draw (0,0) -- (40:1.1);
\draw (0,0) -- (50:1.1);
\node[anchor=west] at (-30:1.1) {\textcolor{aqua}{$k=4$}};
\node[anchor=west] at (15:1.1) {\textcolor{dcyan}{$k=3$}};
\node[anchor=west] at (30:1.1) {\textcolor{dgreen}{$k=2$}};
\node[anchor=west] at (45:1.1) {\textcolor{ddgreen}{$k=1$}};
\begin{scope}[rotate=180]
\fill[fill=dcyan] (0,0) -- (10:1) arc (10:120:1) -- (0,0);
\fill[fill=green!75!black] (0,0) -- (20:1) arc (20:80:1) -- (0,0);
\fill[fill=ddgreen] (0,0) -- (40:1) arc (40:50:1) -- (0,0);
\draw (0,0) -- (10:1.1);
\draw (0,0) -- (120:1.1); 
\draw (0,0) -- (20:1.1);
\draw (0,0) -- (80:1.1) ;
\draw (0,0) -- (40:1.1);
\draw (0,0) -- (50:1.1);
\end{scope}
\draw[->, thick] (0,-1.2) -- (0,1.2) node[right] {$x_2$};
\draw[->, thick] (-1.2,0) -- (1.2,0) node[right] {$x_1$};
\end{tikzpicture}
    \caption{An example of isochronous subsets $\Rs_k$ for a two-dimensional linear PETC system.}
    \label{fig:cones_linear}
\end{figure}

Consider the PETC system \eqref{etc_sys},\eqref{petc_trig_cond} where the plant is a linear system without disturbances
\begin{equation*}
    \dot{\zetav}(t) = \Am\zetav(t) + \Bm\Km\zetav(t_i)), \quad t\in[t_i,t_{i+1}),
\end{equation*}
where $\Am$ and $\Bm$ are the plant matrices and $\Km$ is the controller gain. Suppose further that the triggering function is quadratic, i.e., $\phi(\zetav(t), \zetav(t_i)) = \begin{bsmallmatrix}\zetav(t) \\ \zetav(t_i) \end{bsmallmatrix}\tran
		\!\Qm \begin{bsmallmatrix}\zetav(t) \\ \zetav(t_i)\end{bsmallmatrix}.$
Such quadratic triggering functions include the most common ones in the literature \cite{heemels2013periodic}.
This case has some special properties that allow for tighter abstractions than in the nonlinear case. First, from any sampled state one can determine exactly the state reached after $k$:
\begin{equation}
    \begin{aligned}
        \zetav(t) &= \Mm(kh)\zetav_{\xv}(t_i), \ \forall t \in [t_i, t_{i+1}] \\
        \Mm(t) &\coloneqq \e^{\Am t} + \int_0^{t}\e^{\Am\tau}\d\tau \Bm\Km \eqqcolon \Am_\d(t) + \Bm_\d(t)\Km.
    \end{aligned}
\end{equation}
An \emph{isochronous subset}, which is the part of the state space that triggers at a given inter-sample time $k$, can be implicitly represented by a conjunction of quadratic inequalities \cite{gleizer2020scalable}:
\begin{equation}\label{eq:setq}
	\begin{gathered}
		\Rs_k \coloneqq \Qs_k \setminus \left(\bigcap_{j=1}^{k-1} \Qs_{j}\right) = \Qs_k \cap \bigcap_{j=1}^{k-1} \bar{\Qs}_{j}, \\
		\Qs_k \coloneqq \begin{cases}
			\{\xv \in \R^{n_\zeta} \mid \xv\tran\Nm(hk)\xv > 0\}, & k < k_{\max{}}, \\
			\R^{n_\zeta}, & k = k_{\max{}},
		\end{cases}, \\
		\Nm(hk) \coloneqq \begin{bmatrix}\Mm(hk) \\ \I\end{bmatrix}\tran
		\Qm \begin{bmatrix}\Mm(hk) \\ \I\end{bmatrix}.
	\end{gathered}
\end{equation}
In \eqref{eq:setq}, the set $\Qs_k$ is the set of all points in the state-space that certainly triggered by time $hk$; then the set $\Rs_k$ is the isochronous subset obtained by removing from $\Qs_k$ all points that have triggered before, i.e., $\Qs_{j}, j < k$. The sets $\Rs_k$ can be checked for non-emptiness exactly via nonlinear satisfiability-modulo-theories (SMT) solvers, e.g., Z3 \cite{demoura2008z3}, or approximately through a semi-definite relaxation (SDR) as proposed in \cite{gleizer2020scalable}. The non-empty sets form the finite state-space of a \emph{quotient system} \cite{tabuada2007event}, which gives formally a (exact) simulation relation to the original PETC traffic. Effectively, the sets $\Rs_k$ form a partition of $\R^{n_\zeta}$, and each $\Rs_k$ is a conjunction of finitely many quadratic cones, as depicted in Fig.~\ref{fig:cones_linear}.

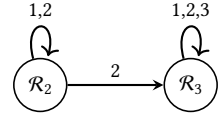
\begin{figure}
    \centering
    \small
      \centering
      \begin{tikzpicture} [node distance = 2cm, on grid, auto, semithick]
        \node (q2) [state] {$\Rs_2$};
        \node (q3) [state, right = of q2] {$\Rs_3$};
        \path [-stealth, thick]
            (q2) edge node {2}   (q3)
            (q2) edge [loop above]  node {1,2} (q2)
            (q3) edge [loop above]  node {1,2,3} (q3);
    \end{tikzpicture}
    \caption{A simple PETC traffic model with only two regions and $k_{\max} = 3.$.} 
    \label{fig:simple_petc_fts}
\end{figure}

The transition relation is as well a problem of checking non-emptiness of an intersection of quadratic inequalities:
\begin{equation}\label{eq:lintran}
    \exists \xv \in \Rs_i : \Mm(hk)\xv \in \Rs_j \implies \Rs_i \edge{k} \Rs_j.
\end{equation}
Similar to the CETC case, the state space of the abstraction is composed of the regions $\Rs_k$; however, the output set is now finite, $\Ys = \{1,2,...,k_{\max}\}$. The transition set is composed of all triplets $\Rs_i \edge{k} \Rs_j$ satisfying \eqref{eq:lintran}; for scheduler synthesis (see Sec.~\ref{sec:scheduler synthesis}), the action set is $\Us = \Ys$, meaning that the scheduler chooses the inter-sample time $kh$, often limited to be smaller than the PETC's inter-sample time $ih$. For verification, we drop the action set and the transitions are $\Rs_i \edge{} \Rs_j \coloneqq \Rs_i \edge{i} \Rs_j$. In both cases, we have effectively a finite transition system (see an example in Fig.~\ref{fig:simple_petc_fts}), in contrast to the CETC case where the inter-sample time can still be chosen on a real interval.

Note that using relaxations for an existence problem such as the one above can lead to spurious transitions. From an abstraction (and simulation) perspective, this is acceptable, as it only adds nondeterminism to the abstraction; any scheduler that works for such an abstraction will work for the original system. Nevertheless, the more one can remove spurious non-determinism, the better, thus using exact solvers such as Z3 is recommended if the dimensions of the system permit. Further tightening can be obtained by using the refinements described in \cite{gleizer2020towards}, where isochronous subsets were extended to \emph{isosequential subsets}, that is, sets $\Rs_{k_1k_2k_3...k_l}$ containing the states whose next $l$ inter-sample times are $hk_1, hk_2, ... hk_l$. This effectivelly chooses the \emph{depth} of the abstraction refinement one wants to use for the traffic model. In ETCetera, the user can set the \lstinline[basicstyle=\ttfamily\small, language=iPython]{depth} value and choose \lstinline[basicstyle=\ttfamily\small, language=iPython]!solver in {'sdr', 'z3'}!; see §\ref{ssec:example_quant} for an example.

\subsection{Scheduler synthesis}\label{sec:scheduler synthesis} 

Once one has constructed traffic models, they can be employed for the synthesis of schedulers. The goal of a scheduler, in our context, is to specify when each of the considered ETC systems should trigger, such that no overlap of their events occur, while still retaining stability of each of the ETC systems (which is guaranteed if always triggering before $\tau(x)$). 

The tool currently considers two methods to synthesize schedulers. The first method relies on the use of UPPAAL. As already indicated in Section 3.1, the general traffic models constructed by ETCetera are equivalent to Timed Automata. Our tool exports the resulting models as timed automata that can be read and manipulated by UPPAAL to synthesize schedulers or check properties of the systems at hand.

The second synthesis method~\cite{Straalen2021} follows from recognizing that the specialized traffic models constructed for PETC are finite transition systems (see Sec.~\ref{ssec:abstraction_linear}). As such this method is only applicable for these traffic models. Furthermore it is assumed that the sampling time for each of the systems is identical.
First, the representation of the traffic models is modified such that it acts on a per-sample basis. Instead of having actions corresponding to the next inter-sample time $\mathcal{U}=\mathcal{Y}=\{1,2,\ldots,k_{max}\}$ as in the models described in Section 3.2, a system can either wait (`$w$') , or trigger (`$t$') at the current sampling time. The result is a new transition system $\hat{\Ss} = (\hat{\Xs}, \hat{\Xs}_0, \hat{\Us}, \hat{\Es}, \hat{\Ys}, \hat{H})$, where:
\begin{itemize}
    \item $\hat{\Xs} := \{T_x \mid x\in\mathcal{X} \} \cup \{W_{x,j} \mid j < H(x),\, x\in\mathcal{X}\}$,
    \item $\hat{\Xs}_0 \subseteq \{T_x \mid x\in\mathcal{X} \}$,
    \item $\hat{\Us} := \{w, t\}$,
    \item $\hat{\Es} := \{(T_x, t, T_{x'}) \mid (x, 1, x') \in \Es\}\allowbreak \cup \{(W_{x,j}, w, W_{x,j+1}) \mid 1 \leq j < H(x)-1\}\allowbreak \cup \{(W_{x,j}, t, T_{x'}) \mid  (x, j+1, x') \in \Es\}$,
    \item $\hat{\Ys} = \{T_1, T, 1, \dots, k_{max}-1 \}$,
    \item $\hat{H}(T_x) = T_1$ if $H(x) = 1$, or $T$ otherwise, and $\hat{H}(W_{i,j}) = H(i)-j$.
\end{itemize}
The outputs indicate either whether the system has just triggered (outputs $T_1$ and $T$)) or the triggering deadline. Some states are given, for technical reasons, the distinct output $T_1$ to capture that from those states the only action that can be taken is `$t$', while on the other states with output $T_i \, (i > 1)$, both `$w$' and `$t$' actions are enabled. In this transition system, triggering after $k$ samples becomes a sequence of $k-1$ `$w$' actions followed by a single `$t$' action.   

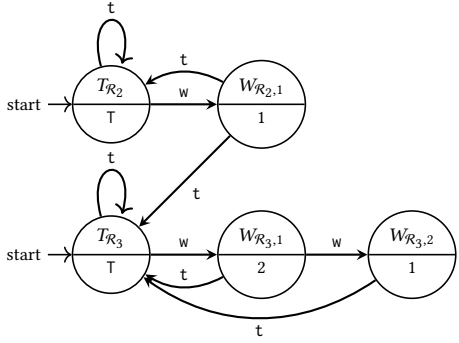
\begin{figure}
    \centering
    \footnotesize
      \centering
      \begin{tikzpicture} [node distance = 2cm, on grid, auto, semithick]
        \node (q00) [state with output, initial] {$T_{\Rs_2}$ \nodepart{lower} \texttt{T}};
        \node (q01) [state with output, right = of q00] {
            $W_{\Rs_2,1}$ \nodepart{lower} 1    };
        \node (q10) [state with output, initial, below = of q00] {$T_{\Rs_3}$ \nodepart{lower} \texttt{T}};
        \node (q11) [state with output, right = of q10] {
            $W_{\Rs_3,1}$ \nodepart{lower} 2};
        \node (q12) [state with output, right = of q11] {
            $W_{\Rs_3,2}$ \nodepart{lower} 1};
            
        \path [-stealth, thick]
            (q00) edge node {\texttt{w}}   (q01)
            (q00) edge [loop above]  node {\texttt{t}} (q00)
            (q01) edge [bend right] node[above] {\texttt{t}}   (q00)
            (q01) edge node {\texttt{t}}   (q10)
            (q10) edge node {\texttt{w}}   (q11)
            (q10) edge [loop above]  node {\texttt{t}} (q10)
            (q11) edge [bend left] node[above] {\texttt{t}}   (q10)
            (q11) edge node {\texttt{w}}   (q12)
            (q12) edge [bend left=35] node {\texttt{t}}   (q10);
    \end{tikzpicture}
    \caption{An example of wait-trigger model $\hat{\Ss}$ for PETC traffic.} 
    \label{fig:twexample}
\end{figure}

Next, all these transition systems (representing different PETC systems) are combined into a larger transition system, through a \textit{parallel composition} allowing each subsystem to behave independently of the rest. Formally, the resulting composition is given by the transition system $\Ss_{comp.} = (\allowbreak \Xs_{comp.}, \allowbreak \Xs_{comp.;0}, \allowbreak \Us_{comp.}, \allowbreak \Es_{comp.}, \allowbreak \Ys_{comp.}, \allowbreak H_{comp.})$, where
\begin{itemize}
    \item $\Xs_{comp.} := \hat{\Xs}_0 \times \dots \times \hat{\Xs}_n$,
    \item $\Xs_{comp.;0} :=  \hat{\Xs}_{0;0} \times \dots \times \hat{\Xs}_{n;0}$,
    \item  $\Us_{comp.} := \hat{\Us}_0 \times \dots \times \hat{\Us}_n$, 
    \item $\Es_{comp.} := \{((x_0, \dots, x_n), (u_0, \dots, u_n), (x'_0, \dots, x'_n) \allowbreak \mid {\forall i \leq n,} \allowbreak (x_i, u_i, x'_i) \in \hat{\Es}_i\},$
    \item  $\Ys_{comp.} := \hat{\Ys}_0 \times \dots \times \hat{\Ys}_n$,
    \item $H_{comp.}(x_0, \dots, x_n) := (\hat{H}_0(x_0), \dots, \hat{H}_n(x_n))$.
\end{itemize}

In this system the behaviour that a scheduler should prevent, i.e. two or more systems triggering simultaneously, is captured  by actions which contain two or more more `$t$'s. Here it is assumed that the fundamental checking period $h$ of each of the PETC loops is the same. Since after triggering a subsystem always lands in a state with output $T$/$T_1$, the scheduler requirement can alternatively be expressed as: `\textit{Avoid states in $\Ss_{comp.}$ whose output contains more than one $T$}'. Creating a controller that is able to avoid these states can be done effectively for finite transition systems by solving a safety game (c.f. Chapter 6 of~\cite{tabuada2009verification}). In this safety game, the maximal fixed point of the operator
\begin{equation}\label{eq:safeop}
	F_W(Z) = \{x \in Z \, \vert \, x \in W \land \exists u \in U(x) : \: \emptyset \neq Post_u(x) \subseteq Z \},
\end{equation}
is found by iterating over its output (starting with $Z_0 := \Xs_{comp.}$). The set $W$ is the \textit{safe set} which in this case is given by:
\begin{equation}
	W = \{(x_1, \dots, x_n) \in \Xs_{comp.} \, \vert \, |\{i \mid \hat{H}_i(x_i) \in \{T_1, T\}\}|  \leq 1 \}.
\end{equation}
The maximal fixed point $Z^* := F_W(Z^*)$ contains states in $\Ss_{comp.}$ (given that $Z^* \neq \emptyset$) from which there exists at least one action guaranteeing that a bad state (a collision of triggering events) can be avoided. The scheduler can subsequently be defined as
\begin{equation}
    U_c(x) = \{u \in \Us_{comp.} \mid \emptyset \neq Post_{u}(x) \subseteq Z \}.
\end{equation}
This is a function that returns a set of \textit{safe} actions given some state $x \in \Xs_{comp.}$, which one can obtain from the last sampled states of each of the control systems. The scheduler keeps track of the current state of each $\hat{\Ss}_i$ and based on this collection of states, $U_c(x)$ provides a safe action, after which each of the states is updated according to $\hat{\Es}_i$. Note that after a transmission, due to the non-determinism of the original traffic model, one needs to rely on the value of the measured plant state to determine the corresponding successor state.

To improve the efficiency and scalability of this approach, two additional techniques are implemented (for details see ~\cite{Straalen2021}):
\begin{enumerate}
    \item \textit{Partitioning and Refinement}: In this approach the size of each $\hat{\Ss}_i$ is reduced by grouping together similar states (\textit{partitioning}). If no scheduler is found for these reduced systems, these blocks are split apart (\textit{refinement}) and the safety game is performed again.
    \item \textit{Binary Decision Diagrams (BDDs)} \cite{bryant1992symbolic}: representing the transition systems with BDDs the manipulation of the transition systems (including partitioning and refinement) can be performed more efficiently, enabling the implementation of the safety-game fixed-point iteration symbolically.
\end{enumerate}

\subsection{Other synthesis and analysis problems}\label{ssec:quant} 

As mentioned in the introduction, traffic abstractions for PETC can also be used for quantitative analysis and other synthesis problems, as recently demonstrated in \cite{gleizer2021hscc, gleizer2021cdc}. For quantitative problems, the transition systems from Def.~\ref{def:system} are equipped with a \emph{weight function} $\gamma : \Es \to \Q,$ mapping edges to rational numbers representing the \emph{weight} or \emph{value} of a transition. These systems are called \emph{weighted transition systems} (WTSs) \cite{chatterjee2010quantitative}. For every run $x_0\edge{u_0}x_1\edge{u_1}x_2...$ of $\Ss$ we can thus compute a sequence of weights $w_0w_1w_2...$, giving rise to the \emph{set of all weight sequences} $\Ws_\Ss$. For ETC, we are interested in computing or maximizing the \emph{smallest average inter-sample time} (SAIST) of a system; hence, the weight of a transition is its inter-sample time. The SAIST of a PETC traffic model $\Ss$ is formally defined as
\begin{equation}
    \mathrm{SAIST}(\Ss) = \inf_{w_0w_1w_2... \in \Ws_\Ss}\liminf_{n\to\infty}\frac{1}{n+1}\sum_{i=0}^n w_i.
\end{equation}
A lower bound to the SAIST of a PETC system can be computed using Karp's algorithm \cite{karp1978characterization, chaturvedi2017note} on the abstraction. In some cases, the abstraction can be a \emph{smallest-average-cycle-equivalent} simulation \cite{gleizer2021hscc, gleizer2021computing}, a property that can be verified for PETC of linear systems: in this case, the SAIST from the abstraction is exact. 

In order to do synthesis, we equip the PETC traffic model with early-triggering samples as in the scheduling problem, and solve a \emph{mean-payoff game} \cite{ehrenfeucht1979positional}, obtaining a lower bound to the optimal sampling strategy \cite{gleizer2021cdc}. The strategy is a function mapping a region $\Rs_i$ onto an inter-sample time choice $k$, becoming effectively a self-triggered control scheme \cite{velasco2003self, anta2008self, mazo2010iss}. The idea behind the approach in \cite{gleizer2021cdc} is to sample early in some regions of the state-space to collect long-term benefits, something that the shortsighted ETC cannot do. The abstractions contain long-term traffic information which allows better strategies to be computed. In \cite{gleizer2021cdc}, the SAIST of a numerical example nearly doubled using this approach.

\subsection{Discussion on computational complexity}
When abstracting general CETC systems, the size of the state-space partition using either gridding or IMs scales exponentially with $n_{\zeta}$. Nonetheless, in the case of linear PETC, the use of isochronous sets removes the dependence on dimensions, and the number of regions is $\mathcal{O}(k_{max}^l)$. The complexity of computing IM-approximations mainly depends on the complexity of the dReal \cite{dreal}, which is exponential on $n_{\zeta}$ for our problem. Generating timing intervals and transitions, in the case of general CETC systems, is $\mathcal{O}(r)C$ and $\mathcal{O}(r^2)C$ respectively, where $r$ is the number of regions and $C$ is the complexity of the reachability analysis operation. For linear PETC, the transition relation computation is $O(k_{max}r^2)C'$, where $C'$ is polynomial in $n_{\zeta}$ when using \lstinline[basicstyle=\ttfamily\small, language=iPython]{solver='sdr'} (see \cite{gleizer2020scalable}) and exponential in $n_{\zeta}$ when using \lstinline[basicstyle=\ttfamily\small, language=iPython]{solver='z3'} (see \cite{gleizer2021computing}). The resulting system $\hat{S}$ for linear PETC has $\mathcal{O}(k_{max}^{l+1})$ states (which is $rk_{max}$) and $\mathcal{O}(k_{max}^{2l+2})$ transitions (which is $rk_{max}^2$). In practice, the number of regions often grows much slower than exponentially with $l$, a phenomenon investigated in \cite{gleizer2022chaos}. Regarding scheduling, using UPPAAL the size of the problem scales exponentially with the number of systems in the network (TA safety games are exponential on the number of clocks). The scheduling technique of Section \ref{sec:scheduler synthesis} is more efficient, exploiting PETC models being already FTSs. Additionally, partitioning can reduce the number of states to $\mathcal{O}(k_{max})$ (and the transitions accordingly to $\mathcal{O}(k_{max}^2)$). However, if the safety game fails, refinement has to be performed, increasing the number of states. Introducing BDDs complicates this analysis, see \cite{Straalen2021} for more details.
It is important to remark, however, that the abstraction and scheduler synthesis processes are performed offline, with the synthesized scheduler’s online implementation being simply a set of if-then-else rules. 

\section{Use of {ETCetera}} 
ETCetera is implemented in Python, and uses third-party tools cvxpy~\cite{diamond2016cvxpy, agrawal2018rewriting}, z3~\cite{demoura2008z3},  dReach~\cite{dreach}, dReal~\cite{dreal} and Flowstar~\cite{flowstar} for the creation of traffic models, UPPAAL Stratego~\cite{uppaal_stratego} for scheduler synthesis and dd~\cite{dd2020} (with bindings to CUDD ~\cite{Somenzi1998Cudd}) for the manipulation of BDDs. To construct traffic models, the user has two possibilities: (i) employ a command line interface, or (ii) employ the utilities provided inside a python script, as illustrated in detail in the examples in Sections~\ref{ssec:example_lin} and ~\ref{sec:Ex:nl}. Analysis and scheduler synthesis can currently only be performed within a Python script, as shown in the examples of Sections~\ref{sec:Ex:sched} and ~\ref{ssec:example_quant}. The computations of all examples below have been conducted on a laptop with an Intel i7-9750H processor and 16 GB of RAM.

\subsection{Command line interface}\label{sec:tool:interface}
If the command line interface is employed, the user simply needs to run the following command:

\begin{lstlisting}[breaklines=true,
    basicstyle=\footnotesize\ttfamily]
$ python etc2traffic.py system_type input_file [options]
\end{lstlisting}

This reads the contents of \textit{input\_file} and generates a traffic model. The contents of \textit{input\_file} depend on the \textit{system\_type} (either \textit{linear PETC} or \textit{general}). When \textit{system\_type} is \textit{linear PETC}, \textit{input\_file} looks like:

\begin{lstlisting}[breaklines=true,
    frame=lines,
    framesep=2mm,
    basicstyle=\footnotesize\ttfamily]
Dynamics : [0 1; -2 3], [0; 1] 
Controller: [1 -4] 
Triggering Sampling Time: 0.01 
Triggering Heartbeat: 0.40
Triggering Condition: [0.95 0 -1 0;0 0.95 0 -1;-1 0 1 0;0 -1 0 1]
\end{lstlisting}
This constructs a specialized PETC traffic model as is described in Section~\ref{ssec:abstraction_linear}. \textit{Dynamics} contains the state evolution matrix ($\Am$) and the input matrix ($\Bm$), and the \textit{Controller} contains the feedback gain ($\Km$). These must all be expressed as matrices. \textit{Triggering Sampling Time} is the checking period $h$ that is considered and \textit{Triggering Heartbeat} is the maximum trigger time $(hk_{max})$. Finally, \textit{Triggering Condition} contains the triggering matrix $\Qm$ as only quadratic triggering conditions are considered in this case.

When \textit{system\_type} is \textit{general}, \textit{input\_file} looks like:

\begin{lstlisting}[breaklines=true,
    frame=lines,
    framesep=2mm,
    basicstyle=\footnotesize\ttfamily]
Hyperbox States : [-2 2], [-2 2]
Grid Points Per Dimension: [3 3]
Dynamics : x1, x1**2*x2 + x2**3 + u1
Controller: -x2 - x1**2*x2 - x2**3
Triggering Condition : e1**2 + e2**2 - 0.01**2
Solver Options : manifolds_times=[0.002 0.0028 0.0038 0.005 0.0065 0.0075], partition_method=manifold, heartbeat=0.021, order_approx=4
\end{lstlisting}
Which constructs a general traffic model as described in Section~\ref{ssec:abstraction_general}. In this case, \textit{Dynamics}, \textit{Controller} and \textit{Triggering Conditions} are specified as symbolic expressions (where $**$ represents exponentiation). State variables are denoted with \lstinline[basicstyle=\ttfamily\small, language=iPython]{x1, x2, ...}, control variables with \lstinline[basicstyle=\ttfamily\small, language=iPython]{u1, u2, ...}, error variables with \lstinline[basicstyle=\ttfamily\small, language=iPython]{e1, e2, ...}, disturbance variables with \lstinline[basicstyle=\ttfamily\small, language=iPython]{d1, d2, ...} and the variable used for homogenization as \lstinline[basicstyle=\ttfamily\small, language=iPython]{w1}. Furthermore, the rectangular region of the state space that is to be considered must be supplied in \textit{Hyperbox States}. Other mandatory arguments depend on: (i) the partition method, (ii) whether the system is homogeneous, (iii) if disturbances are involved:
\begin{itemize}[leftmargin=*]
    \item If \lstinline[basicstyle=\ttfamily\small, language=iPython]{partition_method=grid} or when the system is not homogeneous, \lstinline[basicstyle=\ttfamily\small, language=iPython]{Grid Points Per Dimension} must be specified, which contains the number of regions the state space is divided into for each of the states.
    \item If \lstinline[basicstyle=\ttfamily\small, language=iPython]{partition_method=manifold}, then \lstinline[basicstyle=\ttfamily\small, language=iPython]{manifolds_times} in \lstinline[basicstyle=\ttfamily\small, language=iPython]{Solver}  \lstinline[basicstyle=\ttfamily\small, language=iPython]{Options} must be specified, which will define the Isochronous Manifolds.
    \item If disturbances are present, \lstinline[basicstyle=\ttfamily\small, language=iPython]{Hyperbox Disturbances} must be specified, containing for each disturbance the interval it will lie in.
\end{itemize}
Other optional parameters can be supplied in \lstinline[basicstyle=\ttfamily\small, language=iPython]{Solver Options}. In this example \lstinline[basicstyle=\ttfamily\small, language=iPython]{heartbeat} contains the maximum trigger time and \lstinline[basicstyle=\ttfamily\small, language=iPython]{order_approx} contains the order of approximation of the isochronous manifolds.

In addition to this command line interface, one can construct traffic models via Python scripts as well, as will be shown in Sections~\ref{ssec:example_lin} and ~\ref{sec:Ex:nl}. The command line interface is currently only available for the construction of traffic models, so for scheduling and analysis purposes the procedures in Sections~\ref{sec:Ex:sched} and ~\ref{ssec:example_quant} should be followed.


\subsection{Python interface: linear PETC abstractions}\label{ssec:example_lin}
Consider the linear PETC system:
\begin{equation}\label{eq:ltiTab}
	\begin{split}
		\dot{\zeta}(t) &= \begin{bmatrix} -0.5 & 0 \\ 0 & 3.5 \end{bmatrix}\zeta(t) + \begin{bmatrix} 1 \\ 1\end{bmatrix} u(\zeta(t_i)),  \\
		u(\zeta(t_i)) &= \begin{bmatrix} 1.02 & -5.62   \end{bmatrix}\zeta(t_i),
	\end{split} \quad t\in[t_i,t_{i+1})
\end{equation}
with triggering function $\phi(\zeta(t),\zeta(t_i)) =|\zeta(t)-\zeta(t_i)|^2-(1-\sigma)\left|\zeta(t)\right|^2$, where $\sigma = 0.05$ and $h = 0.01$, $k_{max} = 20$. The PETC traffic model for this system is generated as follows:

\begin{lstlisting}[language=iPython]
import numpy as np

# Define LTI system matrices
A = np.array([[-0.5, 0], [0, 3.5]])
B = np.array([[1],[1]])
K = np.array([[1.02, -5.62]])

# PETC parameters
h = 0.01
kmax = 20
sigma = 0.05

# Triggering condition
Qtrigger = np.block([[(1-sigma)*np.eye(2), -np.eye(2)], [-np.eye(2), np.eye(2)]])

# Construct object representing the PETC system
import ETCetera.Systems as systems

plant = systems.LinearPlant(A, B)
controller = systems.LinearController(K, h)
trigger = systems.LinearQuadraticPETC(plant, controller, kmax=kmax, Qbar=Qtrigger)

# Create Traffic Model
import ETCetera.Abstractions as abstr

traffic = abstr.TrafficModelLinearPETC(trigger)
regions, transitions = traffic.create_abstraction()
\end{lstlisting}

\subsection{Python interface: nonlinear CETC abstractions}\label{sec:Ex:nl}
Consider the nonlinear CETC system:
\begin{equation}
\begin{split}
        \dot{\zeta}(t) &= \begin{bmatrix} \zeta_1(t) \\ \zeta_1^2(t)\zeta_2(t) + \zeta_2^3(t) + u(\zeta(t_i)) \end{bmatrix}, \\
         u(\zeta(t_i)) &= -\zeta_2(t_i) - \zeta_1^2(t)\zeta_2(t) - \zeta_2^3(t),
\end{split}\quad t\in[t_i,t_{i+1})
\end{equation}
together with the triggering condition $\phi(\zeta(t),\zeta(t_i)) =|\zeta(t)-\zeta(t_i)|^2- 0.01^2$. The CETC system may be defined as follows:
\begin{lstlisting}[language=iPython]
import sympy

# Define
state_vector = x1, x2, e1, e2 \
    = sympy.symbols('x1 x2 e1 e2')

# Define controller (in etc form)
u1 = -(x2+e2) - (x1+e1)**2*(x2+e2) - (x2+e2)**3

# Define dynamics
x1dot = -x1
x2dot = x1**2*x2 + x2**3 + u1
dynamics = [x1dot, x2dot, -x1dot, -x2dot]

# Triggering condition & other etc.
trigger = e1 ** 2 + e2 ** 2 - 0.01**2

# State space limits
state_space_limits = [[-2, 2], [-2, 2]]
grid_points_per_dim = [3,3]
\end{lstlisting}

A traffic model for this system is then simply constructed with:
\begin{lstlisting}[language=iPython]
import ETCetera.Abstractions as abstr

# Partition method manifold
traffic = abstr.TrafficModelNonlinearETC(dynamics, trigger, state_vector, state_space_limits=state_space_limits, grid_points_per_dim=grid_points_per_dim, manifolds_times=[0.002, 0.0028, 0.0038, 0.005, 0.0065, 0.0075], partition_method='manifold', heartbeat=0.021, order_approx=4)
regions, transitions = traffic.create_abstraction()
\end{lstlisting}

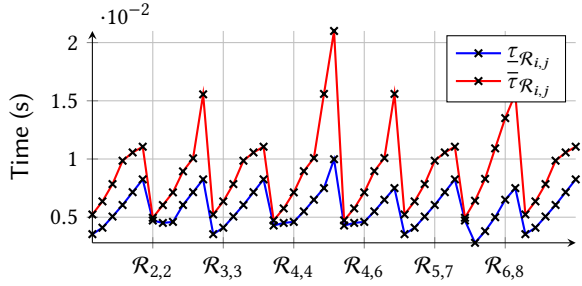
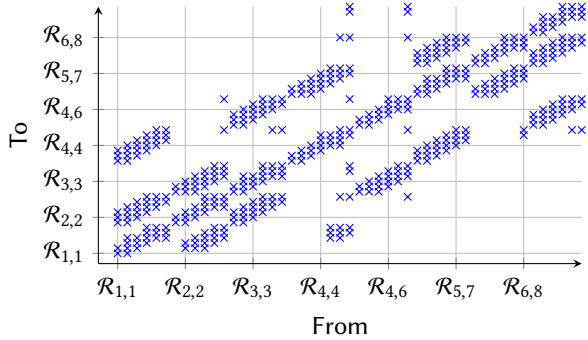
\begin{figure}
    \centering
    \subfloat[Lower and upper bounds of the inter-sample times for each of the regions of the traffic model generated in Example~\ref{sec:Ex:nl}]{\begin{tikzpicture}
	\begin{axis}[height=0.2\textheight, width=0.45\textwidth, my axis style, axis x line=bottom, axis y line=left, ylabel= Time (s), grid=both, xtick={0, 7, 14, 21, 28, 35, 42}, xticklabels={$\mathcal{R}_{1, 1}$, $\mathcal{R}_{2, 2}$,  $\mathcal{R}_{3, 3}$ ,  $\mathcal{R}_{4, 4}$ ,  $\mathcal{R}_{4, 6}$ ,  $\mathcal{R}_{5, 7}$ ,  $\mathcal{R}_{6, 8}$ }]
		\addplot[my linestyle, thick, blue, mark=x, mark options={black}] table[x=r, y=lower, col sep=comma]{Figures/bounds_nonhom_mani.csv};
		\addplot[my linestyle, thick, red, mark=x, mark options={black}] table[x=r, y=upper, col sep=comma]{Figures/bounds_nonhom_mani.csv};
	\legend{$\underline{\tau}_{\mathcal{R}_{i,j}}$, $\overline{\tau}_{\mathcal{R}_{i,j}}$}
	\end{axis}
\end{tikzpicture}}
    \hfill
    \subfloat[Transition between the regions of the traffic model generated in Example~\ref{sec:Ex:nl}]{\begin{tikzpicture}
	\begin{axis}[height=5cm, width=0.45\textwidth, my axis style, axis x line=bottom, axis y line=left, ylabel= To, xlabel=From, grid=both,  only marks, xtick={0, 7, 14, 21, 28, 35, 42}, ytick={0, 7, 14, 21, 28, 35, 42}, ymin=-2, xmin=-2, yticklabels={$\mathcal{R}_{1, 1}$, $\mathcal{R}_{2, 2}$,  $\mathcal{R}_{3, 3}$ ,  $\mathcal{R}_{4, 4}$ ,  $\mathcal{R}_{4, 6}$ ,  $\mathcal{R}_{5, 7}$ ,  $\mathcal{R}_{6, 8}$ }, xticklabels={$\mathcal{R}_{1, 1}$, $\mathcal{R}_{2, 2}$,  $\mathcal{R}_{3, 3}$ ,  $\mathcal{R}_{4, 4}$ ,  $\mathcal{R}_{4, 6}$ ,  $\mathcal{R}_{5, 7}$ ,  $\mathcal{R}_{6, 8}$ }]
		\addplot[color=blue, mark=x] table[x=X, y=Y, col sep=comma]{Figures/tr_nonhom_mani.csv};
	\end{axis}
\end{tikzpicture}}
    \caption{Example \ref{sec:Ex:nl} results.}
    \label{fig:nl_traffic_res}
\end{figure}

Plots of lower and upper bounds on the inter-sample times and transitions for the resulting traffic model can be generated by running the command \lstinline[basicstyle=\ttfamily\small, language=iPython]!traffic.visualize()!, and they are shown in Figure~\ref{fig:nl_traffic_res}. The total CPU time was about 1 hour. 

To add disturbances to the dynamics, the corresponding parameters (\lstinline[basicstyle=\ttfamily\small, language=iPython]{d1,...}) and their domain \lstinline[basicstyle=\ttfamily\small, language=iPython]{dist_param_domain} have to be declared, as well as \lstinline[basicstyle=\ttfamily\small, language=iPython]{partition_method='grid'} has to be specified:

\begin{lstlisting}[language=iPython]
d1 = sympy.symbols('d1')
x1dot = -x1
x2dot = x1**2*x2 + x2**3 + u1 + d1
dynamics = [x1dot, x2dot, -x1dot, -x2dot]
disturbace_limits = [[-0.1, 0.1]]
traffic = abstr.TrafficModelNonlinearETC(dynamics, trigger, state_vector, state_space_limits=state_space_limits, dist_param=(d1,), dist_param_domain=disturbace_limits, order_approx=2, heartbeat=0.022, grid_points_per_dim=[7,8], partition_method='grid', precision_deltas=1e-6)
\end{lstlisting}

\subsection{Scheduler synthesis}\label{sec:Ex:sched}
In this example, the complete workflow of generating a scheduler for two linear PETC systems is described. Consider the two systems~\cite{tabuada2007event,Hetel2011}:

\begin{equation}\label{eq:ltiTabHet}
	\begin{split}
		\dot{\zeta}_1(t) &= \begin{bmatrix} 0 & 1 \\ -2 & 3 \end{bmatrix}\zeta_1(t) + \begin{bmatrix} 0 \\ 1\end{bmatrix} u_1(\zeta_1(t_i)),  \\
		u_1(\zeta_1(t_i)) &= \begin{bmatrix} 1 & -4 \end{bmatrix}\zeta_1(t_i),\\
		\dot{\zeta}_2(t) &= \begin{bmatrix} -0.5 & 0 \\ 0 & 3.5 \end{bmatrix}\zeta_2(t) + \begin{bmatrix} 1 \\ 1\end{bmatrix} u_2(\zeta_2(t_i)),  \\
		u_2(\zeta_2(t_i)) &= \begin{bmatrix} 1.02 & -5.62   \end{bmatrix}\zeta_2(t_i),
	\end{split}\quad t\in[t_i,t_{i+1})
\end{equation}
both with triggering condition $\phi(\zeta(t),\zeta(t_i)) =|\zeta(t)-\zeta(t_i)|^2 - (1-\sigma)\left|\zeta(t)\right|^2$, where $\sigma = 0.05$ and $h = 0.01$, $k_{1,max} = 40$, $k_{2,max} = 20$. These two systems have been defined in files as specified in Section~\ref{sec:tool:interface}. A scheduler can then be synthesized and simulated as follows:

\begin{lstlisting}[language=iPython]
from ETCetera.util import *

# First construct the traffic models
traffic1 = construct_linearPETC_traffic_from_file('examples/linear_2.txt')
traffic2 = construct_linearPETC_traffic_from_file('examples/linear_3.txt')

import ETCetera.Scheduling.fpiter as sched

# Convert traffic models into different form
cl1 = sched.controlloop(traffic1)
cl2 = sched.controlloop(traffic2)

# Construct system and generate scheduler
S = sched.system([cl1, cl2])
Ux, _ = S.generate_safety_scheduler()

# Simulate
S.simulate(x0=[[1,1],[1,-1]], Tmax=5)
\end{lstlisting}

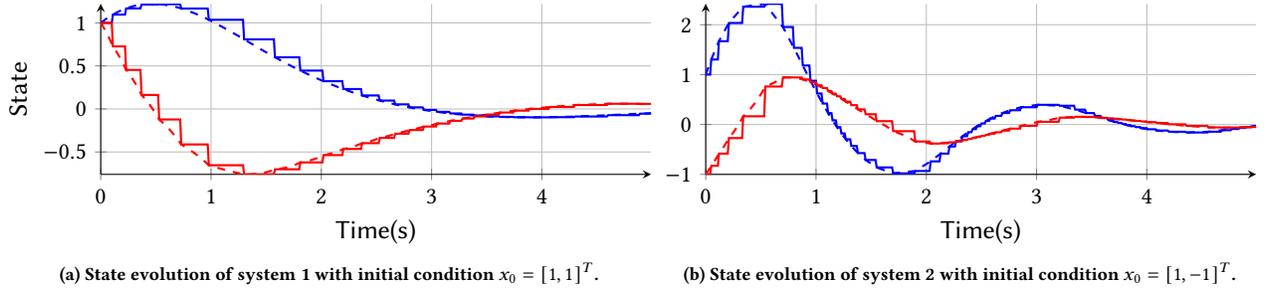
\begin{figure*}
    \centering
    \subfloat[State evolution of system 1 with initial condition $x_0 ={[1,1]}^T$.]{\begin{tikzpicture}
	\begin{axis}[height=0.175\textheight, width=0.5\textwidth, my axis style, axis x line=bottom, axis y line=left, ylabel= State, xlabel=Time(s), grid=both]
		\addplot[my linestyle, thick, dashed, blue] table[y=x11, x=t, col sep=comma]{Figures/sim.csv};
		\addplot[my linestyle, thick, dashed, red] table[y=x12, x=t, col sep=comma]{Figures/sim.csv};
		\addplot[my linestyle, thick, blue] table[y=x1hat1, x=t, col sep=comma]{Figures/sim.csv};
		\addplot[my linestyle, thick, red] table[y=x1hat2, x=t, col sep=comma]{Figures/sim.csv};
	\end{axis}
\end{tikzpicture}}
    \subfloat[State evolution of system 2 with initial condition $x_0 ={[1,-1]}^T$.]{\begin{tikzpicture}
	\begin{axis}[height=0.175\textheight, width=0.5\textwidth, my axis style, axis x line=bottom, axis y line=left, xlabel=Time(s), grid=both]
		\addplot[my linestyle, thick, dashed, blue] table[y=x21, x=t, col sep=comma]{Figures/sim.csv};
		\addplot[my linestyle, thick, dashed, red] table[y=x22, x=t, col sep=comma]{Figures/sim.csv};
		\addplot[my linestyle, thick, blue] table[y=x2hat1, x=t, col sep=comma]{Figures/sim.csv};
		\addplot[my linestyle, thick, red] table[y=x2hat2, x=t, col sep=comma]{Figures/sim.csv};
	\end{axis}
\end{tikzpicture}}
    \caption{Evolution of the systems' states under the scheduler synthesized as detailed in Section~\ref{sec:Ex:sched}. Dashed lines represent the actual state evolution, while the solid lines represent the states available to the controller at each time.}
    \label{fig:sched_sim}
\end{figure*}

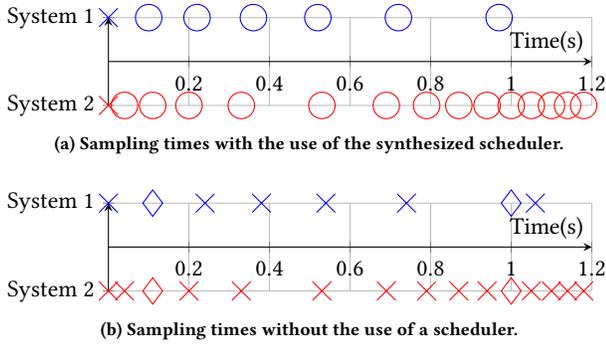
\begin{figure}
    \centering
    \subfloat[Sampling times with the use of the synthesized scheduler.]{\begin{tikzpicture}
	\begin{axis}[height=0.125\textheight, width=0.45\textwidth, my axis style,  only marks, axis x line=middle, axis y line=left,
		yticklabels={System 1, System 2}, ytick={1, -1}, xlabel=Time(s), xmax=1.2]
		\addplot[color=blue, mark=x, mark size = 5pt] table[y=y, x=TT, col sep=comma]{Figures/simTT1.csv};
		\addplot[color=red, mark=x, mark size = 5pt] table[y=y, x=TT, col sep=comma] {Figures/simTT2.csv};
		\addplot[color=blue, mark=o, mark size = 5pt] table[y=y, x=TTe, col sep=comma]{Figures/simTT1.csv};
		\addplot[color=red, mark=o, mark size = 5pt] table[y=y, x=TTe, col sep=comma] {Figures/simTT2.csv};
	\end{axis}
\end{tikzpicture}}
    \hfill
    \subfloat[Sampling times without the use of a scheduler.]{\begin{tikzpicture}
	\begin{axis}[height=0.125\textheight, width=0.45\textwidth, my axis style,  only marks, axis x line=middle, axis y line=left,
		yticklabels={System 1, System 2}, ytick={1, -1}, xlabel=Time(s), xmax=1.2]
		\addplot[color=blue, mark=x, mark size = 5pt] table[y=y, x=TT, col sep=comma]{Figures/sim_refTT1.csv};
		\addplot[color=red, mark=x, mark size = 5pt] table[y=y, x=TT, col sep=comma] {Figures/sim_refTT2.csv};
		\addplot[color=blue, mark=diamond, mark size = 5pt] table[y=y, x=TTc, col sep=comma]{Figures/sim_refTT1.csv};
		\addplot[color=red, mark=diamond, mark size = 5pt] table[y=y, x=TTc, col sep=comma] {Figures/sim_refTT2.csv};
	\end{axis}
\end{tikzpicture}}
    \caption{Sampling times of the first 1.2 seconds of the simulation in Figure~\ref{fig:sched_sim}. For reference, the sampling times when no scheduler is used are shown in (b). Early triggers are denoted with a `$\circ$'. Sampling collisions are represented with a `$\Diamond$'.}
    \label{fig:sched_sim_TT}
\end{figure}

The results of the simulation are shown in Figures~\ref{fig:sched_sim} and ~\ref{fig:sched_sim_TT}. The total CPU time was 10 minutes.
By default, this synthesis approach makes use of BDDs. To make this process more insightful, one can specify \lstinline[basicstyle=\ttfamily\small, language=iPython]{use_bdd=False} in the construction of the control loop. However, this can drastically impact performance of the synthesis process. 

Creating schedulers using UPPAAL follows the same steps, with the exception that additionally a network has to be defined:
\begin{lstlisting}[language=iPython]
import ETCetera.Scheduling.NTGA as sched

cl1 = sched.controlloop(traffic1)
cl2 = sched.controlloop(traffic2)

# Define a network and combine
net = sched.network()
nta = sched.nta(net, [cl1, cl2])

# Generate a strategy using UPPAAL and parse its results
nta.generate_strategy(parse_strategy=True)
\end{lstlisting}

\subsection{Quantitative analysis and synthesis}\label{ssec:example_quant} 

Consider the first system in \eqref{eq:ltiTabHet}. In this example we are interested in obtaining a large average inter-sample time; hence we use the more aggressive Lyapunov-based triggering condition \cite{szymanek2019periodic, gleizer2020towards} of the form
	$ V(\hat\zetav(t)) > -\rho \hat\zetav(t)\tran\Ql\hat\zetav(t), $
	where $\hat\zetav(t) \coloneqq \Am_\d(h)\zetav(t) + \Bm_\d(h)\Km\zetav(t_i)$ is the state prediction after $h$ time units, $V(\xv) = \xv\tran\Pl\xv$, and $0 < \rho < 1$ is the triggering parameter. For the numerical results below, the Lyapunov matrices were taken from \cite{tabuada2007event} as $\Pl = \begin{bsmallmatrix}1 & 0.25 \\ 0.25 & 1\end{bsmallmatrix}, \ \Ql = \begin{bsmallmatrix}0.5 & 0.25 \\ 0.25 & 1.5\end{bsmallmatrix}$, and we set $h = 0.1,$ $\rho=0.8$ and $K = 20$. The triggering condition used is quadratic and can be specified as in §\ref{ssec:example_lin}. The code snippet below (replacing lines 26--27 from the example in §\ref{ssec:example_lin}) shows how to
\begin{enumerate}
    \item construct abstractions to compute the system's smallest average inter-sample time (SAIST), where we flag \lstinline[basicstyle=\ttfamily\small, language=iPython]{etc_only=True} to avoid computing unnecessary early-triggering transitions;
    \item construct a normal abstraction with early-triggering transitions to synthesize a near-optimal sampling strategy;
    \item verify the SAIST of the closed-loop system with the generated strategy.
\end{enumerate}

\begin{lstlisting}[language=iPython, firstline=26, firstnumber=26]
# Best selection for SAIST: solver='z3', etc_only, stop_if_mace_equivalent, smart_mace
traffic = abstr.TrafficModelLinearPETC(trigger, depth=100, etc_only=True, solver='z3', stop_if_mace_equivalent=True, smart_mace=True)
regions, transitions = traffic.create_abstraction()

# Print obtained SAIST and its minimizing cycle
print(f'SAIST is {traffic.limavg}'); print(f'Smallest average cycles: {traffic.cycles}')

# Now create an abstraction to design an improved sampling strategy
traffic_synth = abstr.TrafficModelLinearPETC(trigger, depth=1, etc_only=False, solver='z3', early_trigger_only=True)
strat, strat_tran = traffic_synth.optimize_sampling_strategy()
print(strat) # Inspect strategy

# Compute its SAIST value
traffic2 = abstr.TrafficModelLinearPETC(trigger, depth=100, etc_only=True, solver='z3', stop_if_mace_equivalent=True, smart_mace=True, strategy=strat, strategy_transition=strat_tran)
regions2, transitions2 = traffic2.create_abstraction()
print(f'Optimized SAIST is {traffic2.limavg}')
\end{lstlisting}
This code reproduces part of the results in \cite{gleizer2021cdc}, giving the following output (the numbers are normalized by $h$):
\begin{lstlisting}[breaklines=true,
    frame=lines,
    framesep=2mm,
    basicstyle=\footnotesize\ttfamily]
SAIST is 2.3333333333333335
Smallest average cycles: {(8, 1, 1, 1, 1, 2)}
Optimized SAIST is 5.0
\end{lstlisting}
The algorithm terminated in 5 minutes.
\section{Conclusions and future work}

We have introduced the tool ETCetera enabling the study and manipulation of the inter-sample times generated by event-triggered control systems. The main functionality of the tool is to create abstract \emph{traffic models} from the inter-sample times dynamics implicitly determined by the hybrid dynamics of ETC systems. The tool allows to construct traffic models in the form of timed-automata for general ETC systems, and in the form of finite transition systems in the particular case of PETC systems. The resulting models can be exported for further analysis to UPPAAL, in the case of timed-automata models, or manipulated within the tool (in the case of PETC systems). ETCetera provides algorithmic implementations to synthesize schedulers, compute performance metrics, and design improved sampling strategies for traffic models of PETC systems. The tool is available as an open-source project and continuously being extended with additional functionalities. In particular ongoing and future work is focusing on the following: 
\begin{itemize}[leftmargin=*]
    \item IM-based partitioning for systems with disturbances. Even though the theory developed in \cite{delimpaltadakis2021traffic} addresses this case, this is not yet implemented as Flow* cannot handle ball-segments and dReach cannot handle disturbances. 
    \item In \cite{delimpaltadakis2021traffic} it has been observed that IM-based partitioning, sometimes, results into more transitions, probably because the sets $\Rs_i$ are approximated by ball-segments (which may be a crude approximation). We are working on tackling this issue (with tighter approximations) in future versions of the tool.
    \item Based on the developments of \cite{delimpaltadakis2021stochastic} and \cite{delimpaltadakis2022formal}, extend abstractions to stochastic PETC systems via \emph{interval markov decision processes}, for verification of various performance metrics and for scheduler synthesis with probabilistic safety.
    \item Automate the synthesis of schedulers for PETC loops not sharing the same fundamental sampling time $h$.
    \item Extend the specialized models of linear PETC traffic to perturbed systems, output-feedback systems and CETC.
    \item Enable the simulation of multiple ETC loops under a synthesized scheduler over the network simulator OMNET++ \cite{varga2010omnet++}. 
\end{itemize}
\begin{acks}
The authors thank Dr. Khushraj Madnani for developing the abstraction minimization code and his overall support with scheduler synthesis algorithms, Dr. Cees F. Verdier for assisting in developing the part of the code which generates the IMs, and Dr. Gururaj Maddodi for assisting in the overall architecture of the tool's code. This work is supported by the \grantsponsor{GSERC}{European Research Council}{https://erc.europa.eu/} through the SENTIENT project, Grant No.~\grantnum[https://cordis.europa.eu/project/id/755953]{GSERC}{ERC-2017-STG \#755953}.
\end{acks}

\bibliographystyle{ACM-Reference-Format}
\bibliography{thebib}

%

\end{document}